\documentclass{article}
\begin{document}

\title{Minimal Set of Quantum Postulates and Realistic Interpretation of Quantum Mechanics}
\author{Sumio Wada \\ Institute of Physics \\ The University of Tokyo, Komaba \\ Tokyo, 153-8902, Japan \\ \thanks{Electric address: wada@hep1.c.u-tokyo.ac.jp}}
\date{September, 2009}
\maketitle

\begin{abstract}
We argue that quantum mechanics makes sense without such controversial postulates as the wave function collapse, the quantum probability rule and the observable postulate. We only need the existence of a wave	function as a representation of a state, its dynamical evolution rule and another rather trivial postulate proposed here (Read-Off Postulate). The wave function collapse and the probability rule are replaced with intrinsic properties of the theory, i.e. decoherence and relative frequency, respectively. Interpretation of the wave function as a faithful representation of the reality naturally emerges, although quantum reality has a peculiar feature which distinguishes it from classical reality.

\begin{flushleft}
\end{flushleft}

\end{abstract}

\section{Background}
Interpretation of classical mechanics is simple. The state of a point particle at a given	time t is represented by a position vector $\textbf{r}(t)$. Its time dependence is governed and determined by Newton equation. A principal message of this paper is that interpretation of quantum mechanics is admittedly more complicated but not far more different from that of classical mechanics.  The state is represented by a wave function and its time dependence is governed and determined by Schr\"odinger equation. There is not anymore to be added to these premises besides a rather trivial postulate proposed later.

Most arguments against such a simple interpretation come from two statements which are usually included in a set of postulates/axioms of quantum mechanics, namely the wave function collapse(WFC) and the quantum probability rule(QPR).  Firstly in this section we critically recapitulate these two postulates.

In literatures we can see two different	ways to formulate them. One is an approach based on the state, and the other is the one based on the state operator (the density matrix).  The argument in the state-based approach goes as follows:  Consider a measurement of a certain physical quantity for a certain particle. Suppose that the state of the particle is described as
\begin{equation}
       |\psi> \  = \sum c_\alpha|\psi_\alpha>
\end{equation}
where $|\psi_\alpha>$ are eigenstates of the operator which is associated with the physical quantity we are interested in. Both $|\psi>$ and $|\psi_\alpha>$ are assumed to be normalized.  We also assume that the state $|\psi_\alpha>$ changes to $|\psi'_\alpha>$ after the measurement. Then the linearity of Schr\"odinger equation tells us that $|\psi>$ changes to
\begin{equation}
	     |\psi'> \ = \sum c_\alpha |\psi'_\alpha>
\end{equation}
This state is a superposition of multiple results of the measurement, which looks to be counterintuitive at first sight, and so proponents of WFC and QPR assert that the result of the measurement is not the above, but simply $|\psi'_\alpha>$ for some $\alpha$ with the probability $|c_\alpha|^2$.

However, we may argue that any measurements is a phenomenon caused by a set of atoms and so should be governed by Schr\"odinger equation. Then why should we assume WFC which is clearly inconsistent with Schr\"odinger equation?  Furthermore, it would be regret (at least for	some people) if the fundamental theory of the Nature is not deterministic but probabilistic.

Another approach to WFC and QPR is based on the density matrix $\Xi$, though similar criticism applies to this, too. What corresponds to WFC in this approach is the change of $\Xi$ from a pure state to a mixed state:
\begin{equation}
 \Xi = |\psi><\psi| \ \to \ \Xi = \sum |c_\alpha|^2 |\psi_\alpha> <\psi_\alpha|
\end{equation}
and the right side is interpreted as a statistical ensemble with a weight $|c_\alpha|^2$. Firstly, however, it is difficult to derive this type of changes of $\Xi$ for closed systems, at least in the present formulation of quantum mechanics, and secondly it is not clear how we can reconcile the interpretation of $|\psi_\alpha>$ as a state of a single particle with the interpretation of $\Xi$ as a statistical ensemble.  Note also that a statistical ensemble is based on the existence of many particles but in quantum mechanics there is a well-defined way for the description of many particle states, which is clearly different from the one-particle density matrix.

Dissatisfaction with the WFC/QPR approach also comes from ontological considerations.  Suppose that we measure the position x of the particle.  The representation which corresponds to eq.(1) is
\begin{equation}
       |\psi> = \int dx \ \psi(x) |x>
\end{equation}
Then $|\psi(x)|^2$ is the probability density for the result that the particle is discovered at x. However, what does $|\psi>$ or $\psi(x)$ mean before the measurement is made or when no measurement is made at all? Note that we cannot say that the particle exists at x with probability $|\psi(x)|^2$, because, before the measurement, the states $|x>'s$ are not mutually exclusive but all of them coexist at the same time so that interference is possible. Probability can be assigned only to a set of mutually exclusive events.  In the WFC/QPR approach, only after the measurement has been done, the position of the particle reduces to a certain point and	$|x>'s$ become mutually exclusive.  However, this	means, at least in this approach, that $\psi(x)$ can not be a representation of the reality before the	measurement because an abrupt shrink to a point is against a fundamental principles of physics, i.e. the special relativity.

It is my view, which I hopefully share with many, that a fundamental theory of the Nature should faithfully describe the state of the Nature irrespectively of whether there exist any human beings who make observation or measurement on it. The theory should describes what had happened to the Universe before any living	beings emerged there.  The main purpose of this paper is to argue that quantum mechanics in the present form, when reexamined from a different point of view, can be interpreted to describe the state of the Nature irrespective of whether measurement is made or not, and, furthermore, that it does describe the state of the Nature deterministically.

This means that we assert that the wave	function $\psi(x)$ corresponds to some reality. They are not merely a means to derive information about results of measurements.  Namely we are advocating the realism against the positivism. However, we should also stress that the present form of quantum mechanics forces us to introduce some aspects of the reality which are counterintuitive to us who have been long accustomed to classical reality.

A notable counterintuitive aspect of quantum mechanics is \emph{entanglement}. This means that, in general, properties of a particle can not be defined independently from those of another particle how far apart they are from each other, when the two particles underwent interaction at some earlier stage.  In another words their state can not be expressed in a factorized form. 

A result of measurement is a typical example  of entanglement.  In eqs.(1) and	(2) we only wrote down the state vector of a particle(an observed system), but more rigorous expression is the one in which the state vector of the measurement apparatus is taken into account.  Let $|M_0>$ be	the state of the apparatus before the measurement, while $|M_\alpha>$ be the one after that when the particle	is in the state $|\psi_\alpha>$.  We	also wrote that	the state of a particle after the measurement is $|\psi'_\alpha>$, which means that the state $|\psi_\alpha>|M_0>$	of the combined	particle-apparatus system changes to $|\psi'_\alpha>|M_\alpha>$. When the	initial	state of the particle is a superposition eq.(1), the linearity of  Schr\"odinger equation tells us that the combined state evolves as 
\begin{equation}
  \{\sum c_\alpha |\psi_\alpha>\}|M_0> \ \to\  \sum c_\alpha |\psi'_\alpha>|M_\alpha> 
\end{equation}
if WFC does not	take place.  The right side is an entangled state.	The state of the particle is defined only relative to the state of the apparatus, and vice versa.  We may also add the state $|O>$ of an observer who reads the result registered in the	apparatus($O$ stands for an observer). If we do so,	eq.(5) is rewritten as
\begin{equation}
 \{\sum c_\alpha |\psi_\alpha>\}|M_0>|O_0> \ \to\  \sum c_\alpha |\psi'_\alpha>|M_\alpha>|O_\alpha> 
\end{equation}

In no collapse approach	which we adopt in this paper, no single	state is selected in the right side	of this	expression.  All the terms survive. From the stand point of the	realism	this means that	the reality is superposed. We may also say that the realities coexist in plurality.  This	is obviously counterintuitive but not incompatible with what we actually	observe. Eq.(6) tells us that an observer perceives only a single result $\alpha$ even if we do not invoke WFC.

As can be seen from this argument, basic strategy which we adopt in this paper is one version among	many which have	been inspired by Everett work in ref.1. In a word, these are no collapse approach.  Time evolution of states is	governed only by Schr\"odinger equation.  The	state which is expressed by a single term transforms into superposition as a result of measurement, which Everett called branching.	Nowadays we often call it  branching into many worlds. However,	the original Everett's argument	has been criticized because it was not explained how it is guaranteed that interference of different	terms do not create observable physical	effects.  The controversy on this point	do not seem to have settled yet, but many people argue that decoherence-type mechanism gives us the answer.  I share this opinion and elaborate the argument later in sec.4.

Another	striking feature in Everett's approach is	that he	argued that not	only WFC is unnecessary but	also QPR can be	derived. Instead of probability	he introduced a	quantity called	measure. It was claimed	that the measure of each term in eq.(1)	should be $|c_\alpha|^2$ by some reasoning such as the	additivity. Then suppose that there are	N times	repetition of the identical system $|\psi>$	and that we make measurement for all of them.  For simplicity let there be only two	eigenstates and $|\psi>$ is	written	as 
\begin{equation}
  |\psi>\ = \ a|\alpha>+\ b|\beta> \ \ (|a|^2 + |b|^2 = 1)
\end{equation}
The record registered in the apparatus has a form ${i_1,i_2,\dots,i_N}$ where i's are either $\alpha$ or $\beta$. The measure of a sequence	which has the result $\alpha$ $n$ times and the result $\beta$ $N-n$ times is $|a|^{2n}|b|^{2(N-n)}$. The number of such sequencs is $_NC_n$.  Then a simple algebraic calculation tells	us that, in the	limit of $N \to \infty$, the total measure of all the sequences	whose relative frequency of $\alpha$ is $r(=n/N)$ vanishes unless $r = |a|^2$ (see sec.3). Everett claimued that QPR has been derived. However this argument was criticized (rightfully in my opinion) because the introduction of the measure  essentially amounts to the presupposition of	QPR.

Though his argument may	not be sufficient to derive QPR,  it shed light	on a new aspect of QPR, i.e., the probability from a practical point of view is nothing but	the expectation	on the frequency. In this point	of view	the probability is not a property of a single particle but a property of (strictly speaking) infinite number of particles. This approach to	the probability	is called the frequentism or the frequency approach. 

The frequency approach without the measure was introduced in refs.2 and 3.  First we define a multiparticle state
\begin{equation}
	|\Psi>_N \  = \ |\psi,1>|\psi,2>\dots|\psi,N>
\end{equation}
where all of $|\psi,i>$ are in the identical state eq.(7) with respect to the quantity we are interested in (for	example, the spin of a particle).  Let $P$ be the projection operator such that
\begin{equation}
	P |\alpha> = |\alpha>, \ \ P |\beta> = 0
\end{equation}
and $P_i$ is the projection operator which operates	on $|\Psi>$	in eq.(8) and projects out $|\alpha>$ in i-th part $|\psi,i>$ and acts as an identity operator for other parts.  Then we define the	so-called frequency operator $F_N$
\begin{equation}
     F_N = 1/N \ \sum P_i
\end{equation}
An eigenstate of $F_N$ is a state	which has a fixed number of $|\alpha>$ and its eigenvalue is equal to the	relative frequency of $|\alpha>$ (namely the number of $|\alpha>$ over N). Then it was shown in refs.2 and 3 that, in the limit	of $N \to	\infty$, 
\begin{equation}
     |(F_N - |a|^2)|\Psi>_N |^2  \to 0
\end{equation}
Then the following postulate of quantum	mechanics is invoked.

\begin{quotation}
\textbf{Operator Postulate}: Every observable is associated with a Hermitian operator $O$.	When a system is in	an eigenstate of $O$, then a measurement of $O$ will yields its eigenvalue.
\end{quotation}

If we can claim from eq.(11) that, in the limit $N \to \infty$,	$|\Psi>$ is	an eigenstate of the frequency operator with the eigenvalue	$|a|^2$,	then the above postulate tells us that the relative frequency r	is $|a|^2$, which is nothing but	QPR.  However, we can not assert that eq.(11) is equivalent with the equality
\begin{equation}
       (F_\infty - |a|^2)|\Psi>_\infty = 0
\end{equation}
Then it	was claimed that what we can say from eq.(11) is	no more	than the consistency of QPR.  Some years later, in refs.4 and 5,	more sophisticated definition of the frequency operator	was presented, for which eq.(12)	is valid.  However, because something similar to Everett's measure is used in the definition, it was claimed that the definition and so the predicted probability is	not unique(see ref.6).

The purpose of my previous paper ref.7 was to	show that we can circumvent these criticisms within	the framework of the frequency approach.  A strategy is	that we	do not deal with the frequency operator any longer, but instead examine the state itself directly.  We derived the relative frequency representation of the multiparticle state and examined its behavior.  This was shown to have a $\delta$-function type peak	at the value where there should be.  Then it was discussed what kind of a postulate is required in order to derive the quantum probability rule from this result.

It was also argued in that	paper that avoiding the	frequency operator and making use of the state itself is not a technical issue. This change of strategy is intrinsically related to the fundamental problem of quantum mechanics which we mentioned	above, namely, what kind of the	reality	the wave function represents.  If we regard observation	just as	a  kind of various phenomena	of the Nature, then it should be represented in the	wave function and we should read its result from the wave function. Using the operator as is in the above postulate may be equivalent in most of the	case of observed quantities, but it should not	be regarded as a first principle of quantum mechanics.

These are the background for what we are going to present in this paper.  In the next section we propose a minimal set of postulates of quantum mechanics from the view point of realism. Only three postulates are	required.  WFC,	QPR and the Operator Postulate are no longer adopted. Instead a new postulate (Read-Off Postulate) is introduced to replace the operator postulate and	QPR.  Quantum probability rule is further discussed in sec.3.  Sec.4 is a brief explanation of a dynamical feature of quantum theory which is intended to	replace	WFC.  Sec.5 is devoted to ontological consequence of our new sets of postulates.  We will discuss what should be	regarded as the reality in	quantum	mechanics. This	is the ultimate	purpose	of this	paper and also of most of discussion on these topics.

\section{A Proposed Set of Postulates of Quantum Mechanics}

Before we present our new set of postulates, we briefly review familiar postulates in conventional approaches.  Postulates which have been frequently adopted are, among others,
\begin{flushleft}
(1) A quantum system is	described by a state vector which belongs to a certain Hilbert space. A state vector can be explicitly represented by a (multi-component, in general) wave function.

(2) Time evolution of the wave function	of the state is governed, in the differential form, by Schr\"odinger equation and, in the integral form, by path integral.

(3) Every observable is associated with a Hermitian operator which acts in this Hilbert space and whose eigenvalue corresponds to its possible observed value(Operator Postulate).

(4) Wave function collapse(WFC)

(5) Quantum Probability Rule(QPR)
\end{flushleft}

As for (1), more	explicit statement may	be useful.
\begin{flushleft}
(1') N particle state is described by a wave function of the form
\begin{equation}
	 \psi_1(q_1)\psi_2(q_2) \dots \psi_N(q_N)
\end{equation}
or a linear combination	of such functions($q_i$ stands for a set of variables which specify the configuration of the i-th particle). 
\end{flushleft}

Note the explicit reference to the linear combination. This means the entanglement. Property of each particle can not be defined independently,	how far	apart it is from others. For an example,	when the state is expressed as
\begin{equation}
	\psi_1(q_1)\psi_2(q_2)  + \psi_3(q_1)\psi_4(q_2)
\end{equation}
the first particle is in the state $\psi_1$ when	the second is in $\psi_2$,  while the first is in $\psi_3$ when the second is in $\psi_4$. However, we should also note that mere symmetrization or antisymmetrization is not the entanglement.

Anyway we keep this postulate(1) or (1') along with the second in our new set.  As	we have already emphasized,	we discard WFC and QPR.	 As for	the third (the Operator	Postulate), we claimed in sec.1 that the result of observation	should be read off not by using an operator but from a wave	function itself.  Therefore we also discard the	Operator	Postulate but we need something	new which tells us how to read off information from a wave function.

The new	postulate which	was proposed for this purpose in the previous paper (ref.7, see also refs.8 and 9) is as follows:
\begin{quotation}
\textbf{New Postulate(Read-off Postulate)}:	Suppose	that the norm of the state $\Psi$ can be expressed as 
\begin{equation}
	 \|\Psi\|^2 =	\sum \Psi^\ast \cdot \Psi
\end{equation}
where the summation is over all variables which specify the configuration of the whole system(such as the position variables and spin indices). When a variable is continuous the summation is understood to be an integral. Suppose	that, after some integration and summation, the norm is	expressed by the remaining one summation(integration) over	a certain variable q in the form  
\begin{equation}
  \|\Psi\|^2 = \sum_q \rho(q)\ \ (\ or\ =  \int \rho(q) dq)
\end{equation}
and if $\rho(q) \neq 0$	only when $q=q_0$, then the value of $q$ in this state is $q_0$.	
\end{quotation}

Such $q$ may correspond to a familiar physical quantity such as energy or	angular momentum.  It can also be an index of the two-component wave function, in which case $q$ represents the spin. Note, however, that $q$ does not need to one of the	variables in the summation eq.(15), but can be a variable which arises as a result of change of variables.

We present some examples.

1. When $q$ is a continuous variable, this postulate can be applied when $\rho \propto \delta(q-q_0)$. Then the conclusion that $q =	q_0$ is obvious.

2. Suppose that $\Psi$ is a state of one spinless particle, and can be expanded by a complete set of countable number of orthonormal functions ${f_{qk}}$ as
\begin{equation}
     \Psi = \sum_{ck} c_{qk} f_{qk}
\end{equation}
where q	is a certain discreet quantum number and k represents all the rest.
Then
\begin{equation}
 \|\Psi\|^2 = \sum_{qk} |c_{qk}|^2
\end{equation}
If we define $\rho(q)=\sum_k |c_{qk}|^2$ and $\rho(q)=0$ unless $q=q_0$, then the above postulate tells us that $q=q_0$.

3. If $\rho(q)$ in the above example is nonzero for two values of $q$, say $q_1$ and $q_2$, then we introduce a function $q'(q)$,  which is equal to $q$ besides these two values, and which are equal to another certain single value $q'_0$ when $q=q_1$ or $q_2$. Next we define $\rho'(q')$ as $\rho(q')$ if $q' \neq q'_0$ and, when $q' = q'_0,$
\begin{equation}
 \rho'(q'_0) \equiv \rho(q_1) + \rho(q_2) 
\end{equation}
then the above postulate can be	applied	to $\rho'(q')$ and we can say that $q'=q'_0$, namely $q=q_1$ or $q_2$. This example can be easily extended to the case	when $\rho \neq 0$ for more values, and also when q is a continuous variable	and $\rho(q) \neq 0$ for a	certain	interval of q.

4. Correlation. We are often interested in correlation between two quantum numbers, namely, the value of a certain quantum number when the value of some other quantum number($q$) is fixed. Let $\Psi(q,\ldots)$ be an expression of this state which use $q$ as one of variables. When $q$ is fixed to a certain constant $q_0$, we may apply the above Read-off Postulate to $\Psi(q=q_0,\ldots)$ and examine whether we can get some predictions or not. In the context of many world interpretation, we may say, in certain circumstances, that we examine the state in the worlds where $q=q_0$. We will discuss many world interpretation later in sec.4.

Strictly speaking, as for the prediction about observation of physical quantities, we should examine a combined state of particle and apparatus, which can be expressed abstractly as
\begin{equation}
      |\Psi> = |particle>|apparatus>
\end{equation}
By choosing the position of the pointer in the apparatus state as a variable q in the Postulate, information on the result of observation can be obtained. Whether such observation is possible or not, namely whether a certain variable q is a physical quantity (an observable) or not, depends on details of the theory, such as the structure of Hamiltonian.  This means that observables are determined from the	theory and not vice versa.

One might wonder that our new Postulate is mathematically equivalent to the operator postulate, even if they are different from the ontological viewpoint. In the second example above, we assumed the existence of the complete set of a countable number of orthonormal functions. From it we can construct a hermitian operator whose eigenfunctions are $f_{qk}$ and whose eigenvalues are $q$ (or its function). Therefore we can claim that the operator postulate is satisfied in this case. However, the situation is not so simple when the number of degrees of freedom is infinite. In the present approach the existence of such hermitian operator is not a requirement. We need not bother about constructing such an operator. The mathematical criterion for a physical quantity is lowered.

In summary, our new postulate, which is intended to supersede the operator postulate from the viewpoint of the realism, may give us more flexible framework from a practical point of view because the existence of	a hermitian operator is not required for an observable.

\section{Quantum Probability Rule (QPR)}

In the previous paper(ref.7, see also ref.10), we showed that QPR can be derived in the context of the frequency approach on the basis of our Read-off Postulate above. Firstly, in this section, we recapitulate the argument of the pervious paper by using a simple two-level model, but adopt a slightly different procedure for later convenience.

Suppose that there are N idenitical particles, all of which have the same spin state but exist in regions different from each other.  We write the state of the particle in the i-th region as
\begin{equation}
 |\psi,i> = (a\alpha_i^+ +b\beta_i^+)|0>\ \ (|a|^2+|b|^2= 1)
\end{equation}
where $\alpha_i^+$ and $\beta_i^+$ are (fermionic and normalized) creation operator of the particle in the i-th region $(i=1\sim N)$ with spin up($\alpha$) and spin down($\beta$) respectively.  The spin states are assumed to be identical for all i's and so the coefficients a and b are i-independent.  Then the whole state of N particles can be written as
\begin{equation}
 |\Psi>_N = |\psi,1>|\psi,2>\dots|\psi,N>
\end{equation}
Next we define a N particle state in which the relative frequency of spin up is r, which means that $n(=rN)$ among N particles are in the spin up state($\alpha$) and the rest are in the spin down state($\beta$). Such state which we denote as $|r>_N$ is 
\begin{equation}
 |r>_N = K_N\{\alpha_1^+\dots\alpha_n^+\beta_{n+1}^+\dots\beta_N^+ +perm.\}|0>
\end{equation}
$K_N$ is a constant and
\begin{equation}
     _N<r|r>_N \ =\ K_N^2\ _NC_n =1
\end{equation}
By using this state, $|\Psi>_N$ is expanded as
\begin{equation}
  |\Psi>_N\ = \sum K_N^{-1} a^n b^{N-n}|r>_N
\end{equation}
Its norm is
\begin{equation}
 _N<\Psi|\Psi>_N = \sum |K_N^{-1} a^n b^{N-n}|^2\ _N<r|r>_N
		 = \sum\ _NC_n |a^n b^{N-n}|^2
		 \equiv \sum \rho(N,n)
\end{equation}
Note that the norm is expressed as a summation over one variable n.

Next we consider the $N=\infty$ case. Because
\begin{equation}
	|\Psi>_\infty = |\Psi>_N |\Psi>_{\infty-N}
\end{equation}
for an arbitrary value of N, its norm also has a N dependent expression
\begin{equation}
  _\infty<\Psi|\Psi>_\infty = \sum \rho(N,n)
\end{equation}
Then we	define
\begin{equation}
   \tilde{\rho}(r) = \lim_{N \to \infty} N\rho(N,n)
\end{equation}
$(r=n/N,\ 0<r<1)$, which satisfies	the relation
\begin{equation}
 _\infty<\Psi|\Psi>_\infty\ = \int \tilde{\rho}(r) dr 
\end{equation}
Now we show that the new postulate can be applied to $\tilde{\rho}(r)$. In fact, when N and n are very large, 
\begin{equation}
 N\rho(N,n)\propto\sqrt{N}\exp{\{-N(r-|a|^2)^2/(2|ab|^2)\}}
\end{equation}
which means that
\begin{equation}
 \tilde{\rho}(r) = 0 \ \ unless\ \ r = |a|^2
\end{equation}
Then the postulate tells us that, in the state $|\Psi>_\infty$,
\begin{equation}
 r=|a|^2.
\end{equation}
This result for the two-state	system can be easily extended to a general multistate case and also a continuous variable case, as is shown in ref.7.

It is interesting to see a mathematical reason why we can get the result eq.(33). Suppose that the norm of the state eq.(21) were defined as 
\begin{equation}
 \||\psi,i>\|^2 = |a|^p + |b|^p	= 1
\end{equation}
where p is a certain constant which is not necessarily equal to 2. In this case the expansion of the N particle state eq.(23) is unchanged but
\begin{equation}
 \||r>_N\|^2 = K_N^p\ _NC_n
\end{equation}
and so
\begin{equation}
 \||\Psi>_N\|^2 = \sum |K_N^{-1} a^n b^{N-n}|^p\ \||r>_N\|^2
  = \sum\ _NC_n |a^n b^{N-n}|^p
\end{equation}
from which we get the result
\begin{equation}
 r=|a|^p
\end{equation}
as we derived eq.(33) from eq.(26). 

However, it is difficult to reconcile the normalization condition eq.(34) with the Hilbert space structure of quantum mechanics, unless $p=2$, particularly because this is not basis-independent. Therefore we argue that the origin of QPR in the context of the frequency approach lies in the Hilbert space structure of quantum mechanics.

In the rest of this section, we make some comments on how the above result about QPR is related actual situations of measurements. First note that, if the probability is regarded as the relative frequency of the results of measurements, then we should incorporate some form of measurement processes in the above formulation.

We address this point briefly by taking the spin model as an example. Suppose that there is an apparatus for each region which measures the spin state of the i-th particle there. We write the state of the particle-apparatus system (after the measurement) as
\begin{equation}
		\alpha_i^+|+> \ or \  \beta_i^+|->
\end{equation}
where +/- inside the kets represents the record of the measurement, up(+) or down(-).  We assumed that the state of the particle does not change as a result of measurement just for simplicity.

Then we may write the state of the N-particle N-apparatus system with relative frequency $r$ after measurements	as
\begin{equation}
 |r,A>_N = K_N \{\alpha_1^+\dots\alpha_n^+\beta_{n+1}^+\dots\beta_N^+|+\dots-\dots> +\ perm.\}
\end{equation}
($A$ in $|r,A>_N$ denotes N apparatuses.)  Note that $r$ is not only the relative frequency of the spin up state but also the relative frequency of the record + of the measurement.

The expansion of the N particle-apparatus system $|\Psi,A>_N$ in terms of $|r,A>_N$ is the same as eq.(25)
\begin{equation}
	 |\Psi,A>_N = \sum K_N^{-1} a^n b^{N-n}|r,A>_N
\end{equation}
Therefore we can define $\rho$ and apply the postulate in the same way as before and we get the prediction $r = |a|^2$.

In the present case, $r$ has a clear meaning as the relative frequency of + in the record of measurement. However this is not the end of the story. What we predict is that $r = |a|^2$ in the state $|\Psi,A>_\infty$, but what kind of state is this? $|\Psi,A>_\infty$ is a superposition of various states each of which has a different record of measurements. If we adopt WFC, such superposition is not allowed. One of the states with a definite record should be singled out as a result of the measurement.

In this paper, however, WFC has been discarded. All the states with different records are considered to coexist. If we further assume (on the basis of realism) all the states represent something which exists as reality, then we are led to work in the framework of many world interpretation. In this framework $|\Psi,A>_\infty$ represents a multiverse, i.e. a superposition of universes. In this context our prediction that $r = |a|^2$ is a property of this multiverse.  The multiverse contains many universes and our prediction tells us that the value of the relative frequency r should always be $|a|^2$, namely universes with different $r$ do not coexist in this multiverse.  Such ontological aspects of quantum mechanics will be further discussed in sec.5.

The final issue in this section is about finiteness of N in any actual situations. The postulate can only be applied when N is infinite but in reality we cannot repeat measurement infinite times. However, in a rigorous sense, it is quite natural that there is no prediction when N is finite.  For example, we cannot definitely exclude the case in which all the coins show heads in the toss of, say, a hundred coins.  In the standard probability theory all what we can definitely say about a trial of a hundred time is the relative frequency of the results when a trial of a hundred time is repeated infinite times, from which we can tell how rare/frequent a certain result is. If a very rare result occurs in a single trial(for example, a hundred heads for a hundred coins), we feel uncomfortable, but it is not impossible.

Note that the situation is exactly the same in our approach to quantum probability rule. Let us consider again a particle which has two states $|\alpha>$ and $|\beta>$. Then a system of a hundred of such particles (symmetrized with respect to $|\alpha>$ and $|\beta>$) has 101 states each of which is expressed by $|r>_{100}$ in eq.(23) with $N=100$ and with $n=rN$ from 0 to 100. Then 100 times repetition state $|\Psi>_{100}$ (eq.(25) with $N=100$) can be expanded as 
\begin{equation}
 |\Psi>_{100}=\sum_n K_{100}^{-1}\ a^n b^{100-n}|r>_{100}
\end{equation}
There are 101 terms in the right side, and when $|r>_{100}$ is normalized, 
\begin{equation}
	 K_{100}^{-1} =\ _{100}C_n^{1/2}
\end{equation}
Then our QPR, when applied to the above state with 101 terms tells us that the relative frequency $R(n)$ of the event that $\alpha$ is measured n times among 100 measurements ($n=100r$) when infinite number of a 100-particle measurement are done is equal to the square of the coefficient of $|r>$ in eq.(41), namely,
\begin{equation}
	  R(n) =\ _{100}C_n |a|^{2n} |b|^{2(100-n)}
\end{equation}
which is nothing but the value given by the standard probability theory for an event in which the results $\alpha$ and $\beta$ occur with the probability $|a|^2$ and $|b|^2$, respectively.

\section{Alternative to the Wave Function Collapse(WFC)}
Take Schr\"odinger' cat paradox as an example. In a no collapse approach the state of the nucleus-cat system is expressed symbolically
\begin{equation}
        a|nucleus\ not\ decayed>|cat\ alive> 
               + \ b|nucleus\ decayed>|cat\ dead>
\end{equation}
Superposition of a live cat and a dead cat looks counterintuitive at first sight, but many-worlders claim that the two terms express two independent worlds and, as long as we can not observe any interference effect, there is nothing to prevent us from claiming that worlds actually exist in plurality.

A frequently-cited possible flaw in the above argument is that the above expression is not unique. It can be rewritten, in a more abbreviated notation, as
\begin{equation}
  (|notdecayed>+|decayed>)(a|alive>+b|dead>)\linebreak +(|notdecayed>-|decayed>)(a|alive>-b|dead>)
\end{equation}
More generalized expression is also possible. Then how can we assert that each term in eq(44) represents an independent world while each term in other expression does not?

It is not difficult to present an intuitive argument to prefer eq.(44) over others such as eq.(45). Consider the state
\begin{equation}
  |\psi> = |decayed>|alive>. 
\end{equation}
This is an inconsistent state and so, the amplitude for this state should be zero, i.e.
\begin{equation}
          <\psi|\Psi> = 0
\end{equation}
where $|\Psi>$ is the whole state above(eq.(44) or equivalently eq.(45)). In fact, we can show this is zero from either expression (by using $<decayed|notdecayed>=\linebreak <alive|dead>=0$). Note, however, that each term gives zero in eq.(45) while the two terms add up to zero in eq.(45). This means that, in eq.(44), this amplitude vanishes because of interference. So we cannot claim that each term in eq.(45) represents an independent world.

However, this is not the end of the story. Instead of $|decayed>|alive>$, we may also consider the state
\begin{equation}
     |\psi'> = (|notdecayed>-|decayed>)(b|alive>+a|dead>)
\end{equation}
For this state the amplitude is zero for each term in eq.(45) but is not in eq.(44). Does it mean that there is interference between two terms in eq.(44)? Is there any fundamental difference between eq.(46) and eq.(48)?

For purposes of illustration of this issue, we take another famous example of interference, i.e., a quantum version of the 2-slit experiment. The beam of particles(photons, electrons or even molecules) with identical energy produces a stripe-type pattern on the screen placed behind the slits. This is nothing but interference effect of two states of a single particle, namely the state of the particle which has emerged from the first slit and the coexisting state of the same particle which has emerged from the other. We describe the state of the particle on the screen as
\begin{equation}
          |\psi> = |1> + |2>
\end{equation}
where each term represents the state which evolves from the state which emerged from the first (or second) slit.

Suppose that $|x>$ is the state of a particle when it exists at the point $x$ on the screen. Then the amplitude for this state is
\begin{equation}
          <x|\psi> = <x|1> + <x|2>.
\end{equation}
Its square obviously has cross terms, which means that there is interference.

Next suppose that there is something D (a detector or whatever) between the slits and the screen which registers which slit the particle has passed. The state of this system D after the particle passed the first (or second) slit is denoted as $|D1>$ (or $|D2>$) with $<D1|D2>=0$. Then the state of the combined system when the particle is on the screen is
\begin{equation}
          |\psi,D> = |1>|D1> + |2>|D2>
\end{equation}
In this case the whole state should be specified by the two variables, namely the position of the particle and the state of D. For example, for the combination $(x,D1)$ the amplitude is
\begin{equation}
           <x|<D1|\psi,D> = <x|1>
\end{equation}
Similarly for the combination $(x,D2)$, the amplitude is
\begin{equation}
         <x|<D2|\psi,D> = <x|2>
\end{equation}
These are the amplitudes for different events and so there are no interference for the probability(frequency). The frequency for the event in which the particle reaches at $x$ is the sum of the frequency of $(x,D1)$ and that of $(x,D2)$. (Note that either $|D1>$ or $|D2>$ can be a negative measurement. For the argument below, it is sufficient if $|D1>$ and $|D2>$ are macroscopically different from each other.)

This seems to suggest that the two terms in eq.(51) can be treated as if they are independent, non-interfering worlds. They look mutually exclusive. However, consider the combination $(x,D1-D2)$. The amplitude for this combination is
\begin{equation}
       <x|(<D1|-<D2|)|\psi,D> = <x|1>-<x|2>
\end{equation}
This does seem to mean that there is interference between the two terms in eq.(51). However, the issue is whether a state such as $D1-D2$ is what we should take into account when we examine interference of two worlds. This is the same problem as we encountered in Schr\"odinger's cat, in which the state $|alive>-|dead>$ plays a similar role.

To explicitly observe that the two states D1 and D2 are actually coexisting, we need to see their interference. Interference of the states of a particle, such as $|1>$ and $|2>$ above, is easy to observe because the wave functions overlap. The situation is different in the case of detectors. Suppose that the detector shows its observation by the position of a pointer.  The states of the pointer are narrow wave packets and do not overlap each other when it registers different observation.  This fact itself is not an obstacle to observe interference because the pointer may return to the original position after some time.  However, the detector is not a system with just a single degree of freedom. It consists of practically infinite number of internal particles and also undergoes interaction with particles outside, which we call altogether (internal and external) environment. Even after the pointer returns to the original position, the difference remains in the environment. Moreover, the environment consists of practically infinite number of degrees of freedom and an observer is blind to most of their states (where gobserverhmeans not only a living being but also any system which interacts with the detector). We can not specify most of them and should sum up all the possibilities for their states when we calculate the probability (relative frequency). This means that interference term is multiplied by a practically infinite number of inner product of the states of particles of the environments in D1 and D2, which is practically zero. This is nothing but decoherence due to environment(ref.11). This means that $|D1>-|D2>$ is not a state which any observer can recognize and so we need not (and should not) worry about this type of states when possibility of interference of two worlds is examined.

It might be helpful to formulate this argument in some abstraction level. As an ideal model of a macroscopic state, we consider a system of infinite number of particles with a state of the form
\begin{equation}
      |1>|2>|3>|4>|5>\dots
\end{equation}
i.e. a product state. We consider a set of such states which differ from one another only in a finite number of terms and construct a Hilbert space which is spanned by such states. We call each of such Hilbert spaces a component (ref.4 and 6). States which belong to different components are orthogonal. Moreover, because they differ in an infinite number of terms, we cannot observe their interference because of the same reason as above. 

Let us call two states which belong to different components gmacroscopically differenth. Macroscopically different states do not interfere even when they are superposed. In Schr\"odinger's cat case, $|alive>$ and $|dead>$ are macroscopically different, under the assumption that the cat-environment system consists of an infinite number of degrees of freedom, though this is admittedly an approximation.

We also define a macroscopically entangled state as a state in which multiple states which belong to different components are superposed. $|alive>$ +/- $|dead>$ is a macroscopically entangled state. In a rigorous sense even the state $|dead>$ is macroscopically entangled, which means there are multiple macroscopically different dead cat states, which arise because of difference of the time at which the nucleus decays.

By using these concepts, we can summarize the no-collapse approach. When a microscopic system interacts with a classical/macroscopic object (a measurement device or whatever), in general, there arises a state in which multiple classical/macroscopic states are superposed if no wave function collapse is assumed. Then expand the whole state by using macroscopically non-entangled states (each of them belongs to a single component) as a basis. Then the whole state is interpreted as a superposition of non-interfering many worlds each of which corresponds to each term in this expansion. Admittedly this is an approximate statement because an actual classical/macroscopic object does not consists of an infinite number of degrees of freedom. A state in a single component cannot evolve to a macroscopically entangled state in a rigorous sense. However we know that decoherence often works even when there are merely several particles in environment. When an enormous number (in the order of Avogadro number) of particles exist, it would be quite reasonable to presume that the branching to non-interfering many worlds occurs in a practical sense.

\section{Ontological Aspects}

The principal purpose of this paper is to study whether a realistic and deterministic interpretation of quantum mechanics (at its present form) is possible or not. In order to achieve such an interpretation we cannot retain WFC and QPR in the set of postulates. They are clearly against determinism. It is also undesirable to retain the Operator Postulate, because it is not in line with realism.

In this paper we have discussed how to get along with quantum mechanics without these postulates. As a substitute of these postulate we proposed the Read-Off Postulate in sec.2, as a postulate which composes the minimal requirements for quantum mechanics together with the other two postulates which assert the existence of state vectors (or wave functions) and establish the rule for its dynamical evolution. By using only these three postulates we showed that quantum mechanics makes sense.  

These postulates are formally independent from an ontological interpretation of wave functions, but as was discussed in sec.3 and also in sec.4, justification for our argument is closely related to ontological considerations. In sec.3, for example, we discussed how our prediction on the value of $r$ is physically related to what we usually call the probability. In this argument, we interpreted eq.(40) as a superposition of the worlds each of which has its records on the results of measurement. Also in sec.4, we discussed how wave functions are consistently interpreted, in certain circumstances, as a mutually non-interfering superposition of many worlds. In this sense, our new set of quantum postulates are closely related to such interpretation of quantum mechanics.

Because we discard WFC, every term of the wave function coexists. Whether they are mutually non-interfering or not is a situation-dependent dynamical problem. When they are (practically) non-interfering, we say that they represent many worlds. Realities are superposed in plurality, which is a central idea of our interpretation. This means that, in our interpretation, both determinism and realism are recovered, but reality in quantum mechanics turns out to be quite different from that in classical mechanics.

Interpretation without WFC originated from Everett's work in ref.1. He did not explicitly assert the reality of many worlds. After his work various versions of the no-collapse approach have been proposed. The interpretation which explicitly asserts the reality of many worlds, which we call many world interpretation in a narrow sense, is one of them and this is what the present author is advocating in this paper. This is what appears to be the most natural consequence of our propositions.

As an illustration of our interpretation, let us take a 2-slit experiment of an electron, as an example.  Even when one electron is injected each time, the stripe pattern is generated on the screen behind the slits if the injection is repeated many times. This is a clear evidence that there coexist two states for a single electron, i.e. the one which emerges from the first slit and the other which has emerged from the second.  Admittedly we can observe a trace of the electron only at a single point on the screen in each injection. This is a result of the fact that coexisting states become non-interfering one another after the electron hits the screen.

In general a wave function $\psi(x)$ has a width. The state described by $\psi(x)$ can be	interpreted as a superposition of states in each of which the particle has a fixed position x. In this interpretation, $|\psi(x)|^2$ is regarded as a relative measure of coexistence. Note that we do not attach to this measure any mathematical meaning such as probability. All that we can say from the measure is that a state with the measure zero does not coexist. This is a point in which our argument differs from that of Everett.

What was proved in sec.3 is that $|\psi(x)|^2$ is equal to the probability in the sense of the relative frequency. This does not mean that $|\psi(x)|^2$ has an ontological meaning as a probability. $|\psi(x)|^2$ is a quantity related to a single particle while the probability in the sense of the relative frequency is a quantity related to a system of infinite particles. What was proved is that they are numerically equal.  Ontologically they are manifestly different concepts.

In summary, we proposed a minimal set of quantum postulates which allows a deterministic and realistic interpretation of the theory. All that are described in the state (the wave function) correspond to the reality. As a result the reality exists in plurality, which is the most profound nature of quantum mechanics. Evolution of a state is governed only by a single rule, i.e. Schr\"odinger equation. The wave function collapse is replaced with dynamical branching into practically non-interfering states (worlds), and probability in quantum mechanics is not a property of a state of single particle, but is interpreted as a relative frequency in a infinite-particle system. 

\begin{center}
Acknowledgement
\end{center}
The author is indebted for valuable information to C.Bruce, J.B.Hartle, M.Schlosshauer and A.Shimizu.

\begin{flushleft}
References

1)\ H.Evertt,III: Rev. Mod. Phys. \textbf{29} 454 (1957).

2)\ D.Finkelstein: Transactions of the New York Academy of Sciences \textbf{25} 621 (1963).

3)\ J.B.Hartle: Am.J.Phys. \textbf{36} 704 (1968).

4)\ E.Fahri, J.Goldstone, and S.Gutmann: Ann. Phys. \textbf{192} 368 (1989).

5)\ J.Gutmann: Phys. Rev. A \textbf{52} 3560 (1995).

6)\ C.M.Cave and R.Schack: Ann. Phys. (N.Y.) \textbf{315} 123 (2005).

7)\ S.Wada: J. Phys. Soc. Japan \textbf{76} 094004 (2007). 

8)\ J.B.Hartle: in "Gravitation in Astrophysics" 1986 Cargese NATO Advanced Institute.

9)\ S.Wada: Mod. Phys. Lett. \textbf{A3} 645 (1988)

10)\ Original version in S.Wada: Butsuri \textbf{44} 393 (1989)[in Japanese].

11)\ M.Schlosshauer: Decoherence and the Quantum-to-Classical Transition, Springer-Verlag (2007).
\end{flushleft}

\end{document}